\documentclass{PoS}
\PoS{PoS(HEP2005)102}

\title{Inclusive NC and CC diffraction}

\ShortTitle{Inclusive NC and CC diffraction}

\newcommand{\PO}{I\!\!P}

\author{\speaker{Laurent Schoeffel} \thanks{On behalf of the H1 and ZEUS
collaborations}\\

        CEA Saclay, France\\

        E-mail: \email{schoffel@hep.saclay.cea.fr}}

\abstract{
Recent diffractive structure function measurements in deep inelastic
scattering (DIS) by the H1 and ZEUS
experiments at HERA are reviewed. In neutral current (NC) DIS,
precise measurements have been made of diffraction, $ep \rightarrow eXY$,
the signature of which is a large rapidity gap (LRG)
between the hadronic system $X$ and the proton or one of
its low-mass excited states $Y$.
New results on charged current (CC) DIS
are also presented. They are identified in a CC DIS sample by
a large rapidity gap, corresponding to a diffractive process
$ep \rightarrow \nu XY$ with a $W$ boson exchange.
Moreover, within the framework of QCD hard scattering factorisation
in DIS, all these data provide constraints on the diffractive
parton distribution functions of the proton and predict
a large gluon content. For example, the resulting gluon 
distribution, extracted from the H1 measurements, carries an 
integrated fraction of about $70$~\% of the exchanged momentum.
Note that these functions
are crucial inputs for the calculations of diffractive dijet or
charm production in DIS.
These parton distribution functions 
may also be applicable to hadron-hadron scattering provided that
an additional rapidity gap survival probability is taken into account.
}

\FullConference{International Europhysics Conference on High Energy Physics\\

                 July 21st - 27th 2005\\

                 Lisboa, Portugal}

\begin{document}

\section{Measurements in NC DIS and interpretation}
\begin{figure}
\begin{center}
\includegraphics[width=0.9\textwidth]{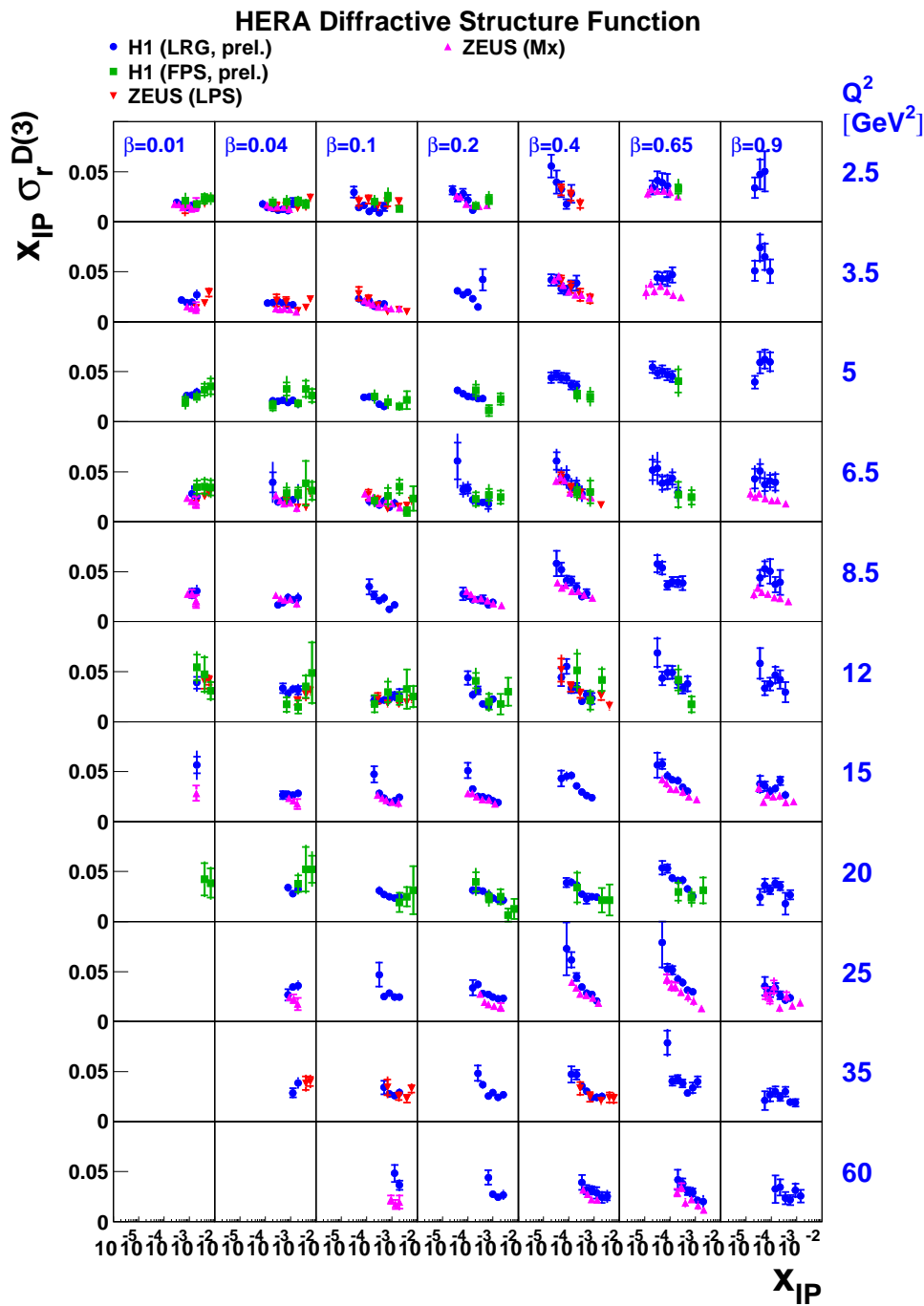}
\caption{Summary of all diffractive DIS data sets at medium $Q^2$,
for $x_{\PO} \sigma_{r}^{D(3)} \simeq x_{\PO} F_2^{D(3)}$ in
$Q^2,\beta$ bins as a function of $x_{\PO}$,
with 
$
\sigma_{r}^{D(3)}
=
\frac{x Q^4}{4 \pi \alpha_{em}^2} \
\frac{1}{(1-y+\frac{y^2}{2})} \
\frac{{\rm d}^3 \sigma_{ep \rightarrow eXY}} { {\rm d} x_{\PO} {\rm d} x {\rm d} Q^2 }
$.
The $Q^2$ and $\beta$ values have been shifted to the H1 bin centers
using small translation factors. Note that all sets have been converted to
the measurement range $M_Y<1.6$ GeV. Thus, ZEUS-$M_X$ data, published for $M_Y<2.3$ GeV,
have been multiplied by
a universal factor $0.77$. The LPS and FPS 
data (leading proton measurements) have been multiplied by factor of $1.1$,
which takes into account the conversion from elastic scattering to
$M_Y<1.6$ GeV.
Relative normalisation uncertainties of order $10$ \% due to these factors
are not shown.}
\label{fig1}
\end{center}
\end{figure}

At low $x$ in DIS at HERA, approximately $10$ \% of the events
are of the type $ep \rightarrow eXp$, where the final state proton
carries more than $95$ \% of the proton beam energy. For these processes,
a photon of virtuality $Q^2$, coupled to the electron (or positron),
undergoes a strong interaction with the proton (or one of its 
low-mass excited states $Y$) to form a hadronic final state
system $X$ (of mass $M_X$) separated by a LRG
from the leading proton, or the system $Y$ (of mass $M_Y$). 
In such a reaction, $ep \rightarrow eXY$,
no net quantum number are exchanged and a 
fraction $x_{\PO}$ of the proton longitudinal momentum is transferred
to the system $X$. In addition, the virtual photon couples to a quark carrying
a fraction $\beta=\frac{x}{x_{\PO}}$ of the exchanged momentum.
Extensive measurements of diffractive DIS cross sections have been made by both
the ZEUS and H1 collaborations~\cite{zeusfwd, zeuslps, h1prel02, h1prel01},
using different techniques~: by requiring explicitely the presence of a LRG,
by exploiting the specific $M_X$ distribution of diffractive events, ot by tagging
the $Y$ system in dedicated detectors. The results
are presented in Fig.~\ref{fig1}.

\begin{figure}
\begin{center}
\includegraphics[width=0.5\textwidth]{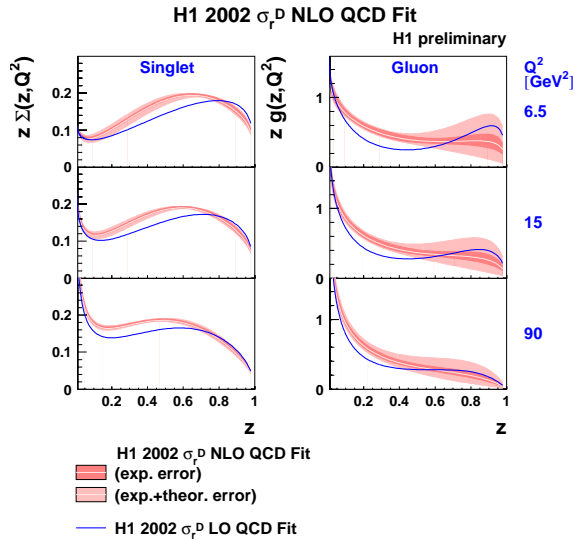}
\caption{
Diffractive parton densities obtained from the QCD fit of the H1 data, normalised
such that the pomeron flux is unity at $x_{\PO}=0.003$. The left hand side shows the
singlet quark distribution ($6 u$ where we assume $u=d=s={\bar u}={\bar d}={\bar s}$).
The right hand side shows the gluon density. 
The red bands show the results of the 
NLO fits, with inner error bands showing the experimental
errors (statistical and systematic) and the other error bands showing the full
uncertainties, including those arising from the theoretical assumptions.
}
\label{fig2}
\end{center}
\end{figure}

Events with the diffractive topology can be analysed in Regge models in terms of
pomeron trajectory exchange between the proton and the virtual photon.
However, the large virtualities $Q^2$ encourage a perturbative QCD treatment.
Then, the diffractive component of DIS is analysed in terms of
parton distribution functions (PDFs). According to the QCD factorisation
theorem~\cite{collins}, these diffractive PDFs (see Fig.~\ref{fig2}) 
will also undergo QCD evolution
as a function of $Q^2$ in the same way as the inclusive proton PDFs.

\section{Measurements in CC DIS}
\begin{figure}
\begin{center}
\includegraphics[width=0.5\textwidth]{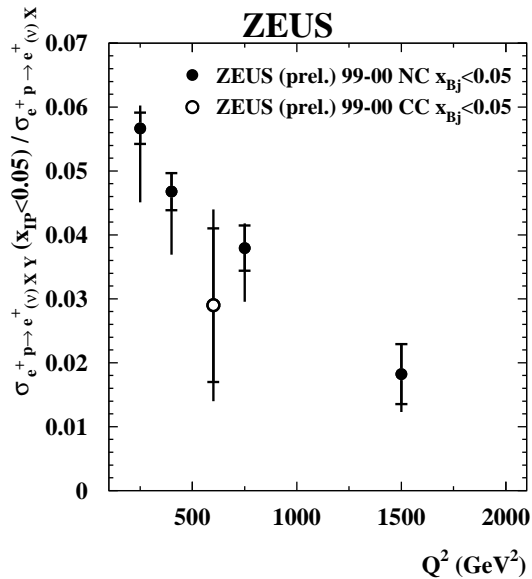}
\caption{
The ratio of the NC LRG cross section $\sigma^{LRG}(ep \rightarrow eXY)$ to
the total cross section $\sigma^{tot}(ep \rightarrow eX)$ as a function
of $Q^2$ is compared to the corresponding ratio for CC.
Similar results are also found by the H1 experiment.
}
\label{fig3}
\end{center}
\end{figure}

Events with a LRG have been observed in $ep$ CC DIS with the
H1 and ZEUS detectors~\cite{zeuscc, h1prel04}. 
Both experiments have presented cross sections
differentially in $Q^2$, $x_{\PO}$ and $\beta$. The ratio of 
diffractive to inclusive NC cross sections has been determined at
$Q^2 > 200$ GeV$^2$ and is compared to that for the CC
process measured in the same kinematic region in Fig.~\ref{fig3} for ZEUS results.
About $2$ to $4$ \%  of the CC events are diffractive, as also measured by
the H1 experiment.
Both collaborations have produced a total CC cross section in the 
kinematic range of the analysis, giving for H1
$\sigma_{CC}^{diff} = 0.42 \pm 0.13 (stat) \pm 0.09 (syst)$~pb,
in good agreement with the RAPGAP Monte-Carlo prediction.

\end{document}